\documentclass{jpconf}
\usepackage[latin1]{inputenc}
\usepackage{graphicx}
\usepackage{amstext}
\usepackage{amssymb}

\DeclareTextFontCommand{\zapf}{\fontencoding{U}\fontfamily{pzd}\selectfont}

\def\ee{$e^+e^-$}               

\def\pp{$pp$}
\def\ppbar{$p\bar p$}

\def\nbar{\bar n}            
\def\Nbar{\bar N}            
\def\nc{{\bar n_c}}          

\def\pt{p\kern -.2pt\lower 4pt\hbox{\tiny T}}    
\def\mt{m\kern -.2pt\lower 4pt\hbox{\tiny T}}    

\def\p0{P_0(\Delta y)}

\def\avg#1{\langle #1 \rangle}  

\def\NF{\mathcal{N}_{\kern -1.9pt f}}
\def\NC{\mathcal{N}_{\kern -1.7pt c}}

\newcommand{\st}[2]{{#1}_{\text{#2}}}
\newcommand{\stt}[3]{{#1}_{{#2},\text{#3}}}

\begin{document}
\title{Clan structure analysis and new physics signals in \pp\
	collisions at LHC}

\author{Alberto Giovannini and Roberto Ugoccioni}

\address{Dipartimento di Fisica Teorica, Università di Torino and
	INFN, Sezione di Torino, via Giuria 1, 10125 Torino, Italy}

\ead{alberto.giovannini@to.infn.it, roberto.ugoccioni@to.infn.it}

\begin{abstract}
The study of possible new physics signals in global event properties
in \pp\ collisions in full phase space and in rapidity intervals
accessible at LHC is presented. The main characteristic is the presence
of an elbow structure in final charged particle MD's in addition to the
shoulder observed at lower c.m.\ energies.
\end{abstract}

\section{Introduction}

The weighted superposition mechanism (WSM) of two properly defined
classes of events (or components) explains some experimental facts
which altogether characterise collective variables properties in high
energy \pp\ collisions and \ee\ annihilation.
In \pp\ collisions the two classes of events are the soft one (without
mini-jets) and the semi-hard one (with mini-jets); in \ee\
annihilation, by using a convenient jet finding algorithm, one
distinguishes between two-jet and three-jet samples of events.
Let us summarise the mentioned experimental facts \cite{ARS:report}:

\noindent 1) shoulder structure in the intermediate $n$-multiplicity
range of the $n$ charged particle multiplicity distribution (MD),
$P_n$, at top c.m.\ energies and in pseudo-rapidity intervals
\cite{ARS:report,Fug,DEL:single};

\noindent 2) quasi-oscillatory behaviour of the ratio of $n$-factorial
cumulants, $K_n$, to $n$-factorial moments, $F_n$, ($H_n = K_n/F_n$ in
the literature) after an initial sharp decrease towards a negative
minimum when plotted as a function of the order $n$ at different c.m.\
energies\cite{ARS:report,hqlett:2,hqlett,NijmegenAG}; 

\noindent 3) forward (F) --- backward (B) multiplicity correlation
strength, $\beta_{FB}$, energy dependence, with
\begin{equation}
	\beta_{FB} = \frac{ \avg{(n_F -\nbar_F) (n_B - \nbar_B)} }{ 
			         \left[\avg{(n_F -\nbar_F)^2}  \avg{(n_B - \nbar_B)^2}
		           \right]^{1/2}} ,  \label{eq:FB.define}
\end{equation}
and $n_F$, $n_B$ the numbers of charged particles lying respectively 
in the forward and backward hemispheres, and $\nbar_F$ and $\nbar_B$
their corresponding average charged multiplicities
\cite{ARS:report,RU:FB,OPAL:FB,RU:FBproblems}.

It should be pointed out that the qualifying assumption of the WSM is
that $P_n$ is described for each class of events in terms of the
Pascal, i.e., negative binomial (NB), MD with the average charged
particle multiplicity, $\nbar$, and $k$ (linked to the variance $D^2
\equiv \avg{n^2} - \nbar^2$ by the relation $k=\nbar^2/(D^2-\nbar)$)
as characteristic parameters, and this fact leads to a sound
description of the experimental data 
\cite{ARS:report,ISR:1+AGCim:3+AGCim:4,UA5:4}.
The NB (Pascal) MD is well known in high energy physics and has been
justified in the framework of QCD.
This approximate regularity has been discovered in the seventies in
all hadronic collisions in full phase-space in the accelerator and ISR
region \cite{ISR:1+AGCim:3+AGCim:4}, 
then extended by UA5 Collaboration \cite{UA5:4} at \ppbar\ collider
energies in the eighties to pseudo-rapidity intervals, and
systematically studied with success by NA22 Collaboration 
\cite{NA22:b1} in \pp\ and
$\pi^{\pm} p$ collisions, by HRS \cite{HRS:1} and Tasso \cite{TASSO} 
Collaborations in \ee\
annihilation, by EMC Collaboration \cite{EMC}
in deep inelastic scattering and by
EHS-RCBC Collaboration \cite{Dengler+Bailly:1988bx}
in proton-nucleus collisions.
These facts led the common wisdom to the conviction that the NB
(Pascal) regularity was an approximate general property of all classes
of collisions \cite{Singapore}, 
which could be interpreted in the framework of clan
structure analysis and understood as a manifestation of a two-step
dynamical process
\cite{AGLVH:0}: to an initial phase in which clan ancestors are
Poissonianly produced, it follows in the second step their decay
according to hadronic showers, each described by a logarithmic MD.
Each clan contains at least one particle (its ancestor) and all
correlations among particles belonging to the same clan are exhausted
within the clan itself.

Clan structure parameters are the average number of clans, $\Nbar$,
and the average number of particles per clan, $\nc$; these new
variables are linked to the standard NB (Pascal) MD parameters,
$\nbar$ and $k$, by the following non-trivial relations:
\begin{equation}
	\Nbar = k \ln \left( 1 + \frac{ \nbar }{ k } \right)
	\qquad\text{and}\qquad
  \nc = \frac{ \nbar }{ \Nbar } .  \label{eq:2}
\end{equation}

Suddenly at the end of the eighties it was found again by UA5
Collaboration that the NB (Pascal) regularity did not survive a more
accurate analysis at top \ppbar\ collider energies \cite{UA5:3}.
The violation of the regularity was confirmed at LEP energies in \ee\
annihilation \cite{DEL:1+DEL:2+DEL:4}. 
Interestingly, the regularity violated in the full sample of events
was rediscovered at a more fundamental level of investigation \cite{Fug}, i.e.,
in the various classes of events contributing to the $n$ charged
particle MD of the total sample, $P_n^{\text{total}}$, which
accordingly was written as follows:
\begin{equation}
	P_n^{\text{total}} = \alpha_1 P_n^{\text{(NB Pascal)}}(\nbar_1,k_1)
    + 
		\alpha_2 P_n^{\text{(NB Pascal)}}(\nbar_2,k_2) , \label{eq:3}
\end{equation}
with $\alpha_1 + \alpha_2 = 1$.
1 and 2 stand respectively for soft and semi-hard in \pp\ collisions
and for 2-jet and 3-jet samples of events in \ee\ annihilation; 
$\alpha_1$ is the weight factor of the first class of events with
respect to the total sample.
Eq.~(\ref{eq:3}) is one essential ingredient of the WSM.
Although our attention will be focused in the following on \pp\
collisions, the WSM is quite general and has been applied successfully
also to \ee\ annihilation.
A remark should be added at this point.
For a correct description of the FB multiplicity correlation strength, 
$\beta_{FB}$, energy dependence in \pp\ collisions, clans of the same
kind of those originally defined on purely statistical grounds are
demanded \cite{RU:FB}.
A result which raises intriguing questions on the real existence of
clans themselves as physically observable quantities \cite{RU:clanmass}.
In conclusion, clan concept seems more close to the real world than a
purely statistical concept.

Accordingly, we decided to examine more carefully clan behaviour in
the possible scenarios obtained by extrapolating the weighted
superposition mechanism from the GeV to the TeV energy domain
\cite{combo:prd+combo:eta}.
Our search was based on the knowledge of the GeV energy region.
Three scenarios were discussed \cite{Ugoccioni:these}.
Following CDF findings at Fermilab it has been assumed that the soft
component satisfies KNO scaling in all scenarios, i.e., $\st{k}{soft}$
remains constant throughout all the explored TeV region.

KNO scaling was also assumed for the semi-hard component in the first
scenario but being KNO scaling behaviour disfavoured by CDF data
\cite{CDF:soft-hard} 
between 630 GeV and 1.8 TeV (and it is unlikely to be verified at
higher c.m.\ energies) this possibility was considered quite extreme
and not realistic.
The semi-hard component is assumed to violate strongly KNO scaling in
the second scenario ($\st{k}{semi-hard}^{-1}$ increases with c.m.\
energy almost linearly in $\ln s$.) 
In the third scenario KNO scaling violation is a QCD inspired one,
i.e.,  $\st{k}{semi-hard}^{-1}$ increases with c.m.\ energy as
$a - b/\sqrt{\ln s}$.

The interest is on the semi-hard component behaviour in the scenario
with strong KNO scaling violation (the second one) and in the QCD
inspired scenario (the third one) and in their clan structure
analysis.
The results of this search are given in Table~\ref{tab:I}.
Common feature of both scenarios is the decrease of the average number
of clans, $\st{\Nbar}{semi-hard}$, and the corresponding increase of
the average number of particles per clan, $\stt{\nbar}{c}{semi-hard}$
as the c.m.\ energy increases from 900 GeV to 14 TeV. The effect is
more pronounced in the second than in the third scenario.
It seems that Van der Waals-like cohesive forces are at works among
clans. Somehow, clan aggregation is occurring and accordingly particle
population density per clan is expected to become larger as c.m.\
energy increases.
This result leads to the following question (See Ref.~\cite{RU:NewPhysics}):
when will clan aggregation in the semi-hard component be maximal?
Of course when $\st{\Nbar}{semi-hard}$ is approximately equal to one
unit, i.e., when between the two parameters of the NB (Pascal) MD,
$\st{\nbar}{semi-hard}$ and $\st{k}{semi-hard}$, the following relation holds:
\begin{equation}
	\st{\nbar}{semi-hard} = \st{k}{semi-hard}\left(
	  e^{1/\st{k}{semi-hard}} - 1 \right) .  \label{eq:4}
\end{equation}
Following the natural decrease of $\st{\Nbar}{semi-hard}$ as the c.m.\
energy increases, relation (\ref{eq:4}) will be reached at a too high
energy to be significant (see Figure \ref{fig:1}).
Notice that $\st{\Nbar}{semi-hard}$ in the first scenario contrary to
the other two is growing with c.m.\ energy: a consequence of 
$\st{k}{semi-hard} \simeq \text{constant}$ and the full control of
$\st{\Nbar}{semi-hard}$ by $\st{\nbar}{semi-hard}$.
As already stated, the study of this scenario has been neglected on
the basis of CDF findings.

\newcommand{\ohmy}[2]{\hspace*{0.1cm}#1\hspace*{7cm}#2\hspace*{4.5cm}~}
\begin{figure}
  \begin{center}
		\small\ohmy{$\Nbar$}{$\nc$}\\
  \mbox{\includegraphics[width=0.9\textwidth]{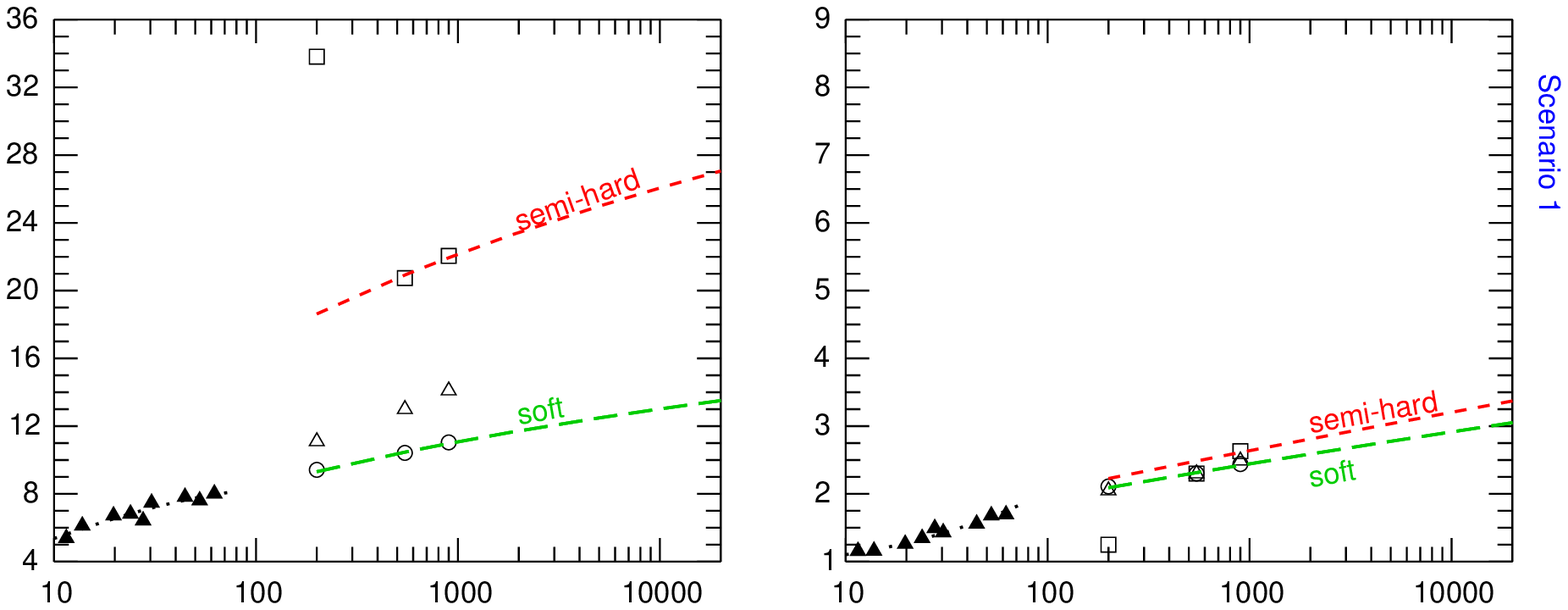}}
  \mbox{\includegraphics[width=0.9\textwidth]{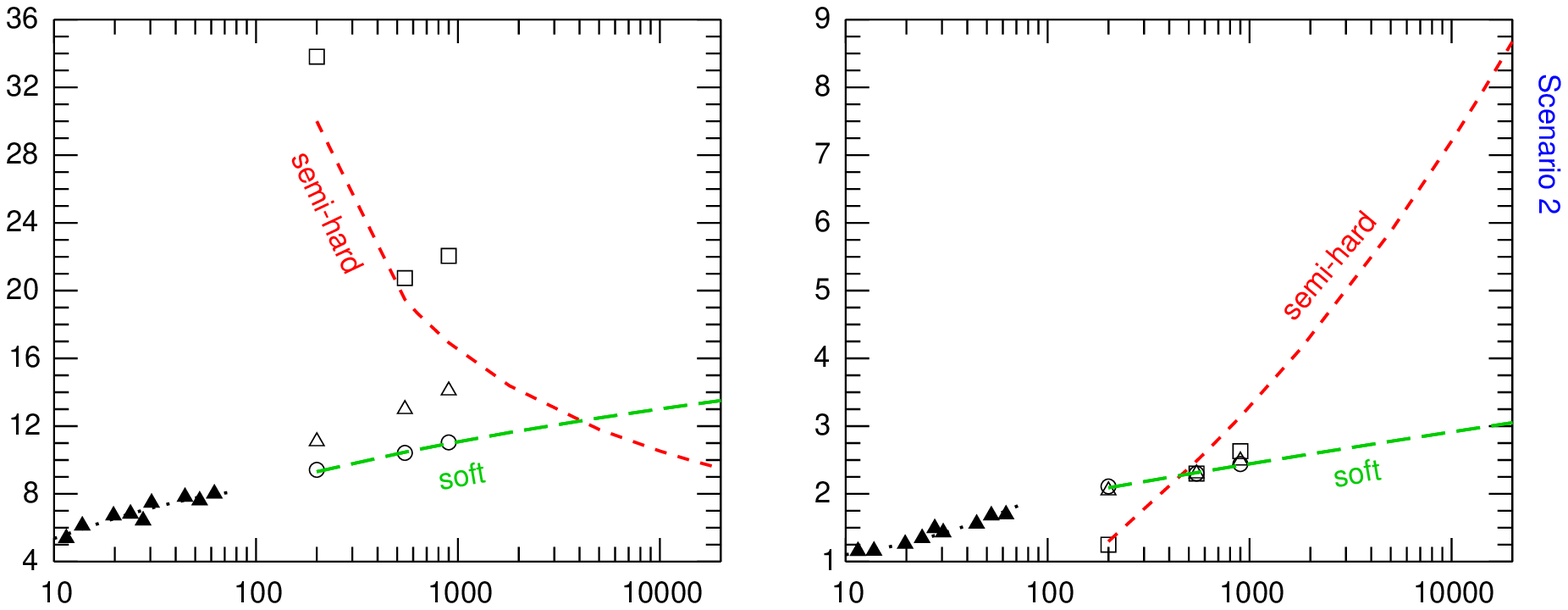}}
  \mbox{\includegraphics[width=0.9\textwidth]{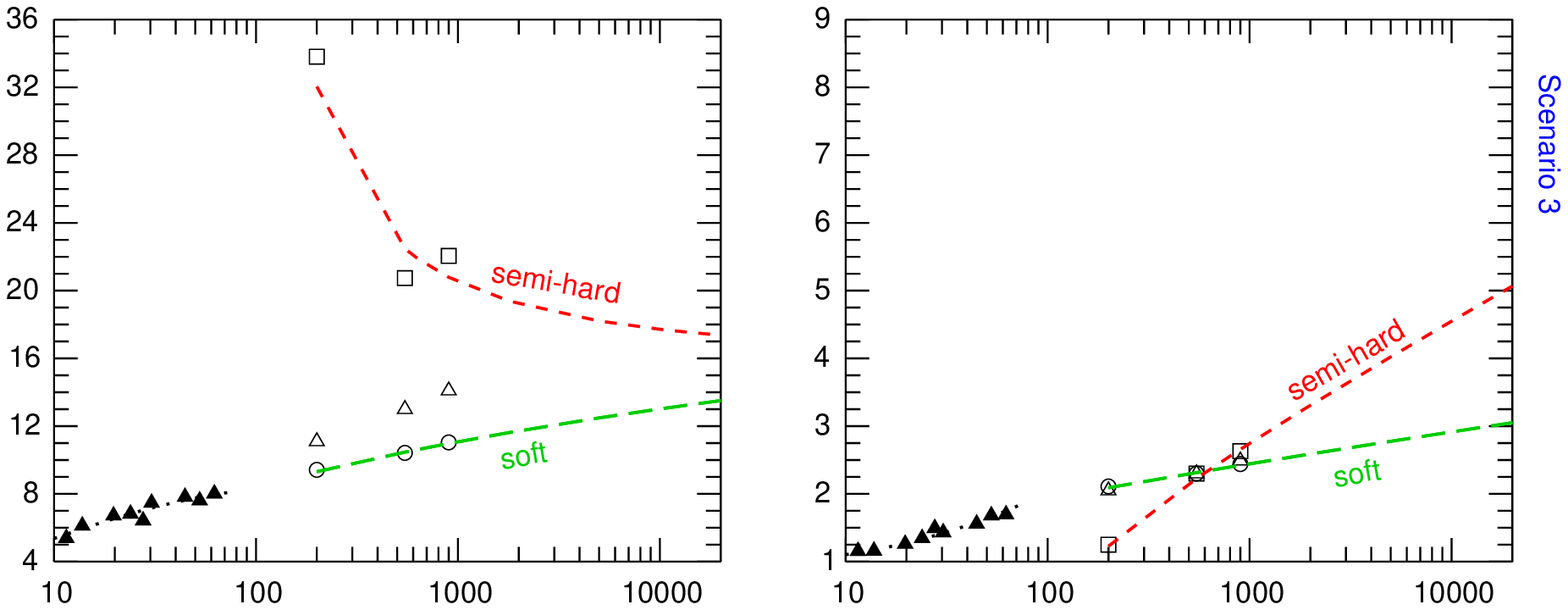}}\\[-0.1cm]
		\ohmy{}{~~~~~~~~~~~~~~\small\textsf{c.m.\ energy (GeV)}}
  \end{center}
  \caption[Extrapolations to TeV: clan parameters]{Clan parameters $\Nbar$
(panels in the left columns) and $\nc$ (panels in the right column)
are plotted for the scenarios described in the text vs.\ c.m.\ energy
(from top to bottom: first row: scenario 1; second row: scenario 2;
third row: scenario 3).
The figures shows experimental data (filled triangles) from
ISR and SPS colliders, the UA5 analysis with two NB(Pascal) MD's of SPS data
(circles: soft component; squares: semi-hard component), together with
our extrapolations (lines: dotted: total distribution; dashed: soft
component; short-dashed: semi-hard component)}\label{fig:1}
\end{figure}

This remark suggests to ask a new question: under which conditions the
decrease of the average number of clans to one unit could be
extrapolated to 14 TeV?
Assuming that these conditions are verified at 14 TeV, are they
related to asymptotic properties of the semi-hard component or are
they the
benchmark of a new class of events, of an effective third component to
be added to the soft and semi-hard ones?
It should be pointed out that the onset of a third class of events in
terms of a second shoulder in $P_n$ vs $n$ is suggested in minimum
bias events in full phase-space by Monte Carlo
calculations with Pythia version 6.210, with default parameters but using
a double Gaussian matter distribution (model 4).

\begin{table}
\caption{Variation of the average number of clans,
	$\st{\Nbar}{semi-hard}$,  and of the
	average number of particles per clan, 
	$\stt{\nbar}{c}{semi-hard}$, between 900 and 14 TeV,
	for the semi-hard component in scenarios II and III.}\label{tab:I}
\begin{center}
  \lineup
	\begin{tabular}{lllll}
		\br
		& \multicolumn{2}{c}{\small $\st{\Nbar}{semi-hard}$}&
		  \multicolumn{2}{c}{\small $\stt{\nbar}{c}{semi-hard}$}\\
		& \small  900 GeV & \small 14 TeV & 
    \small 900 GeV & \small   14 TeV\\
		\mr
		scenario II    & 23 & 11 & 2.5 & 7\\
		scenario III   & 22 & 18 & 2.6 & 5\\
		\br
	\end{tabular}
\end{center}
\end{table}

Coming to the first part of the above question, since one is forced to
exclude that a sudden decrease of the average number of clans to one
unit could be anticipated in the semi-hard component at 14 TeV c.m.\
energy (it would imply heavy discontinuities in
$\st{\nbar}{semi-hard}$ and $\st{k}{semi-hard}$ general behaviours, a
fact which is quite unlikely to occur), it is proposed to consider
relation (\ref{eq:4}) as the benchmark of a new class of events.

Therefore the claim is that
\begin{equation}
	\st{\nbar}{III} = \st{k}{III}\left(
	  e^{1/\st{k}{III}} - 1 \right)   \label{eq:4prime}	
\end{equation}
identifies a new class of events (whose onset was foreseen by the
aforementioned Pythia Monte Carlo calculations),
but with on the average only one clan.
Let us examine now the variation domains of the two new parameters
$\st{\nbar}{III}$ and $\st{k}{III}$ and their influence on the
structure of the MD.
Since $\st{k}{III}$ parameter increases quickly as the c.m.\ energy
increases and just the opposite happens to $\st{\nbar}{III}$, at 14
TeV one should  expect that $\st{\nbar}{III} \gg \st{k}{III}$.
This condition implies that the NB (Pascal) MD of the third component
becomes a gamma MD.
In addition, being $\st{\Nbar}{III}$ reduced to one unit, one should
expect that single clan MD should be quite well approximated by a
logarithmic MD, according to the general rule of clan structure
analysis.
This requests implies that  $\st{k}{III} \to 0$.
The gamma MD for the single clan for $\st{k}{III} < 1$ is indeed a
log-convex distribution, which for $\st{k}{III} \ll 1$ and close to
zero is well approximated by the wanted logarithmic MD.
Altogether, the above conditions
\begin{equation*}
	\st{\nbar}{III} \gg \st{k}{III} \qquad
  \st{k}{III} \ll 1 ~~\text{and}~~ \simeq 0
\end{equation*}
clarify in terms of standard NB (Pascal) MD parameters the deep
meaning of relation (\ref{eq:4prime}) for $\st{\Nbar}{III}\simeq 1$.

\begin{figure}
  \begin{center}
  \includegraphics[width=0.7\textwidth]{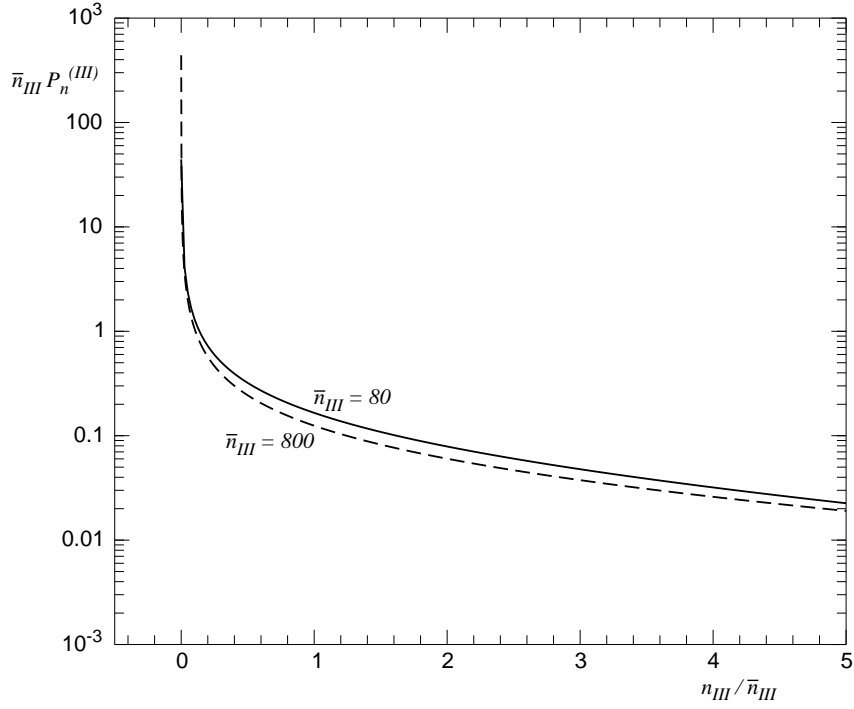}
  \end{center}
  \caption{Multiplicity distributions in KNO form for two values of
		$\st{\nbar}{III}$, with the respective values of $\st{k}{III}$ 
    (0.1611 for $\st{\nbar}{III}=80$
		and 0.1128 for $\st{\nbar}{III}=800$)
    obtained from Eq.~(\ref{eq:4prime}), 
		i.e., requiring $\st{\Nbar}{III}=1$.}\label{fig:2}
\end{figure}

\section{Main properties of the new class of events}

\subsection{$P_n^{\text{(III)}}$, the $n$ charged particle MD}

Coming to the $n$ charged particle MD of the third class of events,
$P_n^{\text{(III)}}$, the plot in Figure~\ref{fig:2} of
$\st{\nbar}{III} P_n^{\text{(III)}}$ vs $\st{n}{III}/\st{\nbar}{III}$
reveals for $\st{n}{III}/\st{\nbar}{III} < 1$ quite large values of 
$\st{\nbar}{III} P_n^{\text{(III)}}$ (events with low multiplicity
with respect to $\st{\nbar}{III}$ are more numerous); a result which
should be compared with the behaviour of $\st{\nbar}{III}
P_n^{\text{(III)}}$ in the region $\st{n}{III}/\st{\nbar}{III} > 1$:
here events with high multiplicity with respect to $\st{\nbar}{III}$
are less probable although they extend for very large multiplicities.

\subsection{Two-particle correlations}

Two-particle correlations of the new component,
\begin{equation}
	\frac{ \st{\nbar}{III}^2}{ \st{k}{III}} = 
	   \int C_2^{\text{(III)}} (\eta_1,\eta_2) d \eta_1 d \eta_2  \gg 
		 \frac{ \st{\nbar}{semi-hard}^2}{ \st{k}{semi-hard}},
\end{equation}
are much larger than two-particle correlations of the semi-hard
component.

\subsection{$n$-factorial cumulant moments, $K_n^{\text{(III)}}$}

Since $\st{k}{III}^{-1}$ controls $n$-factorial cumulant moments
behaviour at any order $n$ for the NB (Pascal) MD and for its limits
in $\st{\nbar}{III}$ and $\st{k}{III}$ parameters,
$K_n^{\text{(III)}}$ is expected to be much larger than
$K_n^{\text{(semi-hard)}}$.

\subsection{Saturation of FB multiplicity correlation strength,
	$\beta_{FB}$}

In $\st{b}{III} \equiv \st{\nbar}{III}/(\st{\nbar}{III} +
\st{k}{III})$, being $\st{\nbar}{III} \gg \st{k}{III}$, with
$\st{k}{III} \to 0$ one should have $\st{b}{III} \to 1$ and since
for $\st{\Nbar}{III} \to 1$ the corresponding leakage
controlling FB multiplicity correlations close to its maximum (leakage
parameter close to 1/2) one gets:
\begin{equation}
   \stt{\beta}{FB}{III} =  \frac{ 2 \st{b}{III} \st{p}{III} 
		 ( 1- \st{p}{III}  ) }{ 1  - 2  \st{b}{III}  
		 \st{p}{III} (1-\st{p}{III}) }   \to 1,
\end{equation}
i.e., $\stt{\beta}{FB}{III}$ saturates and FB multiplicity
correlations in the third component are much stronger than in the
semi-hard class of events.
An indication, in view of the extremely high virtuality and hardness
of these events, of a huge colour exchange process at parton level of
which strong FB multiplicity correlations are presumably the hadronic
signature. 

\bigskip
In conclusion, in this framework one should expect to see at 14 TeV in
\pp\ collisions three classes of events each one described by NB
(Pascal) MD or by its limiting values:

\renewcommand{\labelenumi}{\roman{enumi})}
\begin{enumerate}
	\item the class of soft events with $\st{k}{soft}$ constant as the
	c.m.\ energy increases;
\item the class of semi-hard events with $\st{k}{semi-hard}$ which
	decreases as the c.m.\ energy increases (the class of events of the
	first scenario has been excluded by CDF findings);
\item the class of hard events with $\st{\nbar}{III} \gg \st{k}{III}$
	and $0 \lesssim \st{k}{III} \ll 1$, i.e., $\st{\Nbar}{III} \simeq 1$
\end{enumerate}

\begin{figure}
  \begin{center}
  \includegraphics[width=0.57\textwidth]{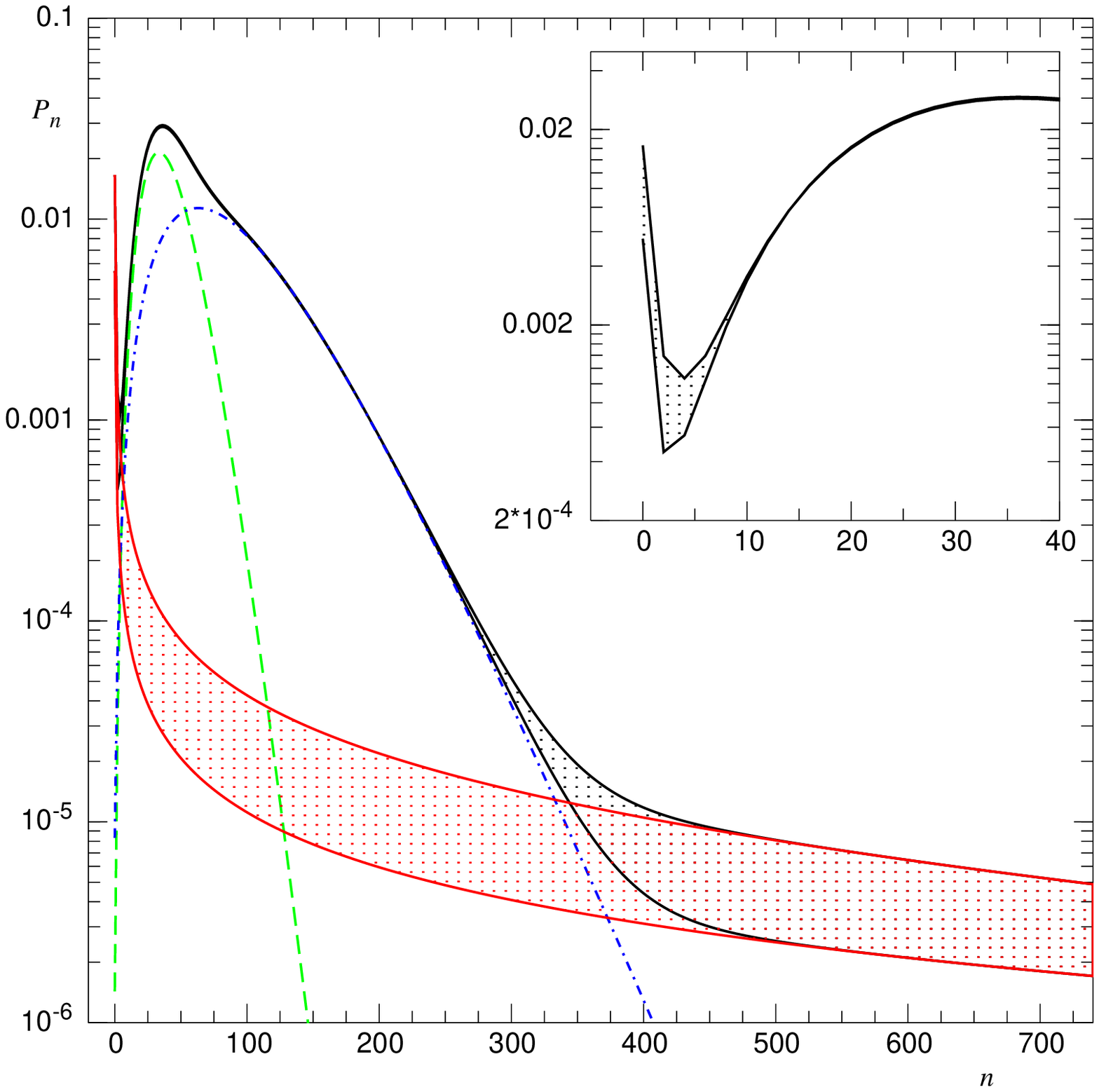}
  \includegraphics[width=0.57\textwidth]{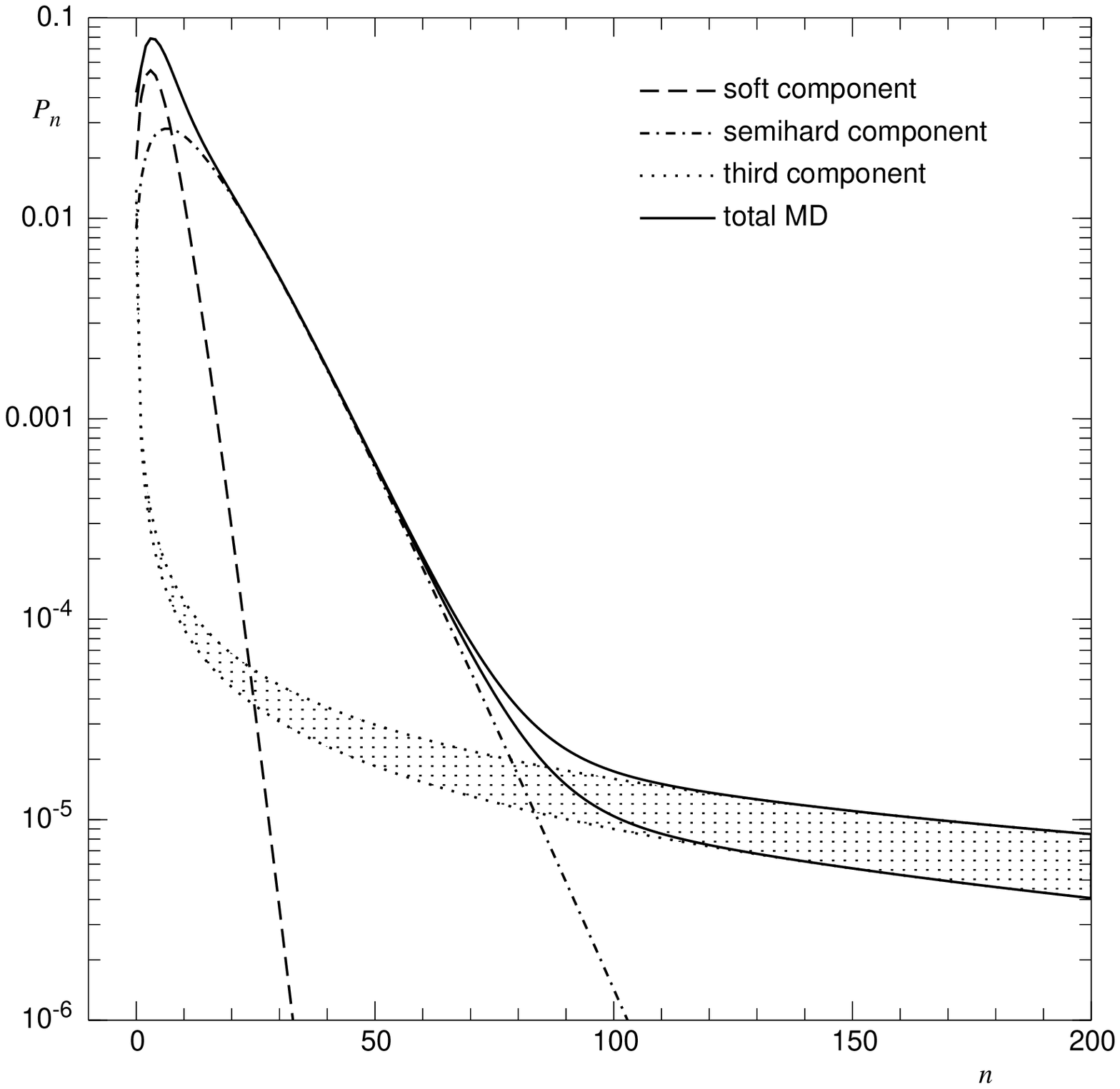}
  \end{center}
  \caption[MD at 14 TeV with elbow]{$n$ charged particle
        multiplicity distribution $P_n^{(\text{total})}$ 
				(solid line) expected at 14 TeV 
				in full phase-space (top panel) and in $|\eta|<0.9$
				(bottom panel) in presence
        of a third (maybe hard) component with
        $\Nbar_{\text{III}}=1$, showing one 
        shoulder structure and one `elbow' structure; the three 
        components are also shown: soft (dashed line), semi-hard
        (dash-dotted line) and the third one (dotted band).}\label{fig:3}
\end{figure}

The total $n$ charged particle MD $P_n^{\text{total}}$ should
therefore be written as follows:
\begin{eqnarray}
	P_n^{(\text{total})} &=& \st{\alpha}{soft} 
	P_n^{\text{(NB Pascal)}} (\st{\nbar}{soft},\st{k}{soft})
	\nonumber\\ &+& 
	\st{\alpha}{semi-hard} 
	 P_n^{\text{(NB Pascal)}}(\st{\nbar}{semi-hard},\st{k}{semi-hard})
	 \nonumber\\ &+&
   \st{\alpha}{III} 
	 P_n^{\text{(NB Pascal)}}(\st{\nbar}{III},\st{k}{III}) ,
   \label{eq:6}
\end{eqnarray}
with $\st{\alpha}{soft} +\st{\alpha}{semi-hard} + \st{\alpha}{III}
=1$, where 
$\st{\alpha}{soft}$, $\st{\alpha}{semi-hard}$ and  $\st{\alpha}{III}$
are the weight factors of the three classes of events with respect to
the total sample of events (See Figure~\ref{fig:3}).

\begin{table}
  \caption{Parameters of the three components at 14 TeV in full
  phase-space.}\label{tab:II}
	\begin{center}
		\lineup
  \begin{tabular}{llllll}
		\br
		  FPS\vphantom{\LARGE H} &    \%   &  $\nbar$  &  $k$ &  $\Nbar$ &  $\nc$\\
      \mr
		  soft    &      41  &   \040        &    7.0     &       13.3    &
		  \0\03.0\\
			semi-hard  &    57  &   \087      &     3.7     &      11.8&
		  \0\07.4\\
			third      &    \02  &  460       &    0.12    &     \01     &
      460\\
			\br
  \end{tabular}
	\end{center}
\end{table}

\begin{table}
  \caption{Parameters of the three components at 14 TeV in the
  pseudo-rapidity interval $|\eta|<0.9$.}\label{tab:III}
	\begin{center}
		\lineup
  \begin{tabular}{llllll}
      \br
		  $|\eta|<0.9$\vphantom{\LARGE H} &    \%   &   $\nbar$   &   $k$
		  &  $\Nbar$ &  $\nc$\\ 
      \mr
		  soft    &41 &    \0\04.9   &  3.4   &   3.0   &   \0\01.6\\
			semi-hard  & 57 &    \014  &  2.0   &   4.2   &   \0\03.4\\
			third (1) & \02  &   \040  &  0.06  &   0.37  &  109\\
			third (2) & \02  &   460  &  0.12  &   1     & 460\\
			\br
  \end{tabular}
	\end{center}
\end{table}

Assuming that at 14 TeV the third class of events is 2\% of the total
sample of events and that $\st{k}{III} \simeq 0.12$ and extrapolating 
$\st{\alpha}{soft}$ and $\st{\alpha}{semi-hard}$ from their behaviour
in the GeV energy range \cite{Ugoccioni:these} one gets in full
phase-space (FPS) the numbers in Table~\ref{tab:II}  (notice that
small variations of $\st{k}{III}$ below 0.12 in Equation
(\ref{eq:4prime}) give $\st{\nbar}{III} \gg 460$.)

In the pseudo-rapidity interval $|\eta| < 0.9$, assuming  (1) that the clan
is spread over all the phase-space or (2) concentrated in $|\eta| <
0.9$ one gets the results in Table~\ref{tab:III}.

\section{Conclusions}

The reduction of $\st{\Nbar}{semi-hard}$ in \pp\ collisions with the
increase of the c.m.\ energy in the TeV energy region (second and
third scenarios discussed in \cite{Ugoccioni:these}) led us to
postulate a third class of hard events to be added to the soft and
semi-hard ones, whose benchmark is $\st{\Nbar}{III} \simeq 1$, i.e.,
$\st{\nbar}{III} \gg \st{k}{III}$ and
$\st{k}{III} \ll 1$ with $\st{k}{III} \simeq 0$.

The main properties of this new class of events are discussed and
predictions at LHC are presented.

The extension of this search to nucleus-nucleus collisions is under
investigation.

\newpage
\section*{References}

\end{document}